\documentclass[prc,nofootinbib,showpacs,12pt]{revtex4}
\usepackage{amsmath,amssymb,graphicx,mathrsfs}
\usepackage{dcolumn}
\usepackage{bm}
\usepackage{feynmf}
\newcommand{\la}{\langle}
\newcommand{\ra}{\rangle}

\newcommand{\GamL}{\Gamma_\Lambda^\mu}

\newcommand{\psie}{\psi_\p}
\newcommand{\psiep}{\bar\psi_{\p'}}
\newcommand{\A}{\alpha}
\newcommand{\B}{\beta}

\newcommand{\p}{{\bf p}}
\newcommand{\kk}{{\bf k}}
\newcommand{\Lam}{\Lambda}
\newcommand{\bLam}{\bar{\Lambda}}
\newcommand{\GL}{\Gamma_\Lambda^\mu}
\newcommand{\OL}{O_\Lambda^\mu}
\newcommand{\OLb}{O_{\bar\Lambda}^\mu}
\newcommand{\VL}{V_\Lambda}
\newcommand{\be}{\begin{equation}}
\newcommand{\ee}{\end{equation}}
\newcommand{\nn}{\nonumber\\}
\newcommand{\eqn}[1]{\label{#1}}
\newcommand{\eq}[1]{Eq.~(\ref{#1})}
\newcommand{\eqs}[1]{Eqs.~(\ref{#1})}

\newcommand{\bpsi}{\bar{\psi}}

\begin{document}
\title{On the Wilsonian renormalization group equation for nuclear current operators}

\author{A. N.  Kvinikhidze}\email{sasha_kvinikhidze@hotmail.com}
\affiliation{A.\ Razmadze Mathematical Institute, Georgian Academy of Sciences, Aleksidze Str.1, Tbilisi 0193, Georgia}
\affiliation{Department of Physics, Flinders University, Bedford Park, SA 5042, Australia}
\author{B. Blankleider}\email{boris.blankleider@flinders.edu.au}
\affiliation{Department of Physics, Flinders University, Bedford Park, SA 5042, Australia}
\date{\today}

\begin{abstract}
We present the solution to the recently derived Wilsonian renormalization group (RG) equation for nuclear current operators. In order to eliminate the present ambiguity in the RG equation itself, we introduce a new condition specifying the cutoff independence of the five point Green function corresponding to the two-body propagator with current operator insertion. The resulting  effective current operator is then shown to obey a modified Ward-Takahashi identity which differs from the usual one, but that nevertheless leads to current conservation. 

\end{abstract}

\maketitle

\section{Introduction}
The use of the Wilsonian renormalization group (RG) method \cite{Wilson:1973jj,Polchinski:1983gv,Morris:1998da} to impose a cutoff $\Lambda$ on the momenta of virtual states is
an important tool for studying various aspects of nuclear effective field theory (EFT) \cite{Birse:1998dk,Birse:2005um,Harada:2005tw,Harada:2006cw,Harada:2007ua,Nakamura:2004ek}.  In this context it has mostly been used to study the strong interactions of nonrelativistic two-nucleon
systems where the central starting point is the RG equation for the two-body effective potential $\VL$ \cite{Birse:1998dk}. Recently, however, Nakamura and Ando (NA) have 
extended the scope of such studies by deriving the RG equation for the two-body effective current operator $\OL$ \cite{Nakamura:2006hc}. The main purpose of the present paper is to present the unambiguous solution to this equation. As our solution differs from the one given by NA, we have endeavored to give a detailed account of both the solution and the RG equation itself. In particular, we present an {\em off-shell}
cutoff-independence condition that leads {\em necessarily} to NA's RG equation. By contrast, NA derived their equation as only a {\em sufficient} condition for an {\em on-shell}
cutoff-independence condition. In this way we eliminate the consequent ambiguity of NA's RG equation. 
 We also eliminate the possibility of any missed subtleties in our solution of the RG equation by obtaining the solution in two different ways: (i) by verifying it through direct substitution into the RG equation, and (ii)  by deriving it explicitly from the equation defining the reduced space current operator. Lastly, we examine the question of current conservation for the derived effective current operator $\OL$. We find that even in the best case where $\OL$ is obtained (via the RG equation) from the full field-theoretic current operator $O^\mu_\infty$ that satisfies the usual Ward-Takahashi (WT) identity \cite{Bentz:1986nq}, $\OL$ will not obey this  identity (it will instead satisfy a modified WT identity).  Nevertheless, the same operator $\OL$ is shown to conserve current in matrix elements. 
 
\section{Solution to the current operator RG equation}
We consider a nonrelativistic two-body system for which the RG method is used to
introduce a  momentum cutoff of $\Lam$. For this purpose it is convenient to use the projection operators \cite{Nakamura:2006hc}
\begin{align}
 \eta &=\int\frac{d^3{\bf k}}{(2\pi)^3}\,|{\bf k}\ra\la{\bf k}|\,\theta(\Lambda-k), \\
\lambda &=\int\frac{d^3{\bf k}}{(2\pi)^3}\,|{\bf k}\ra\la{\bf k}|\,\theta(\bar\Lambda-k)\theta(k-\Lambda),
\end{align}
where $\bLam>\Lam$. The RG equation for the reduced space effective potential $\VL$ \cite{Birse:1998dk} can then be written as
\be
\frac{\partial V_\Lambda}{\partial \Lambda} = V_\Lambda G_0 \frac{\partial \lambda}{\partial \Lambda} V_\Lambda   \eqn{VRG}
\ee
where $G_0=(E-H_0)^{-1}$ is the two-body free propagator and
 \be
 \frac{\partial \lambda}{\partial \Lambda}=-\int\frac{d^3{\bf k}}{(2\pi)^3}\,|{\bf k}\ra\la {\bf k}|\,\delta(k-\Lambda).
 \ee
\eq{VRG} can be derived from the reduced space Lippmann-Schwinger equation
\be
T=V_\Lambda+V_\Lambda\eta G_0 T \eqn{LSE}
\ee
 by using the fact that the off-shell scattering amplitude, $T$, does not depend on $\Lambda$. 
 
 Although \eq{VRG} has been used  to study nuclear EFT
\cite{Birse:1998dk,Birse:2005um,Harada:2005tw,Harada:2006cw,Harada:2007ua,Nakamura:2004ek}, such investigations have been limited to the purely hadronic sector. However, 
in a recent work NA have 
extended the scope of such studies by deriving the corresponding RG equation for the reduced space effective current operator $\OL$  \cite{Nakamura:2006hc}. Writing this current operator as
\be
\OL = \eta \GL \eta,
\ee
the RG equation derived by NA can be expressed as
\begin{align}
 \frac{\partial   \GL }{\partial \Lambda}&=\VL G_0 \frac{\partial \lambda}{\partial \Lambda}\GL
 +\GL \frac{\partial \lambda}{\partial \Lambda}G_0 \VL . \eqn{RGE}
 \end{align}
Here we provide the solution to \eq{RGE}, noting that the solution  given  in Refs.\ \cite{Nakamura:2006hc,Nakamura:2006zp,Nakamura:2007qm} differs from ours. We find, unambiguously, that 
\begin{eqnarray}\label{our}
\OL&=& \eta(1-V_{\bar\Lambda}G_0\lambda)^{-1}\OLb
 (1-\lambda G_0V_{\bar\Lambda})^{-1}\eta
 \nn
 &=& \eta\left[1+V_{\bar\Lambda}\lambda G_0(1-V_{\bar\Lambda}G_0\lambda)^{-1}\right]
 \OLb
 \left[1+(1- \lambda G_0V_{\bar\Lambda})^{-1}G_0 \lambda V_{\bar\Lambda}\right]\eta
 \nn
 &=& \eta\left[1+V_{\bar\Lambda}\lambda (E'-H_0-V_{\bar\Lambda}\lambda)^{-1}\right]
 \OLb
 \left[1+(E-H_0- \lambda V_{\bar\Lambda})^{-1} \lambda V_{\bar\Lambda}\right]\eta
 \end{eqnarray}
 where
$\OL$ is given at $\Lambda=\bar\Lambda$ by its starting value $\OLb$, and $V_{\bar\Lambda}$ is the two-body interaction defined in the model space with the cutoff ${\bar\Lambda}$. The last line of \eq{our} has been written in a form that is most easily compared with Refs.\ \cite{Nakamura:2006hc,Nakamura:2006zp,Nakamura:2007qm}. To prove \eq{our}, we show explicitly that the corresponding $\GL$, 
\be
\GL=(1-V_{\bar\Lambda}G_0\lambda)^{-1}\OLb
 (1-\lambda G_0V_{\bar\Lambda})^{-1},   \eqn{sln}
 \ee
 satisfies \eq{RGE}. 
 
We first use \eqs{VLP} to write \eq{sln} as
\be
 \GL= (1+V_\Lambda G_0\lambda)\OLb
 (1+\lambda G_0V_\Lambda), 
\ee
and therefore
\be
 \frac{\partial   \GL  }{\partial \Lambda}=\frac{\partial V_\Lambda G_0\lambda}{\partial \Lambda} \OLb
 (1+\lambda G_0V_\Lambda)+(1+V_\Lambda G_0\lambda)\OLb
 \frac{\partial \lambda G_0V_\Lambda}{\partial \Lambda}.  \eqn{tmp}
 \ee
 The use of the RG equation for $\VL$, \eq{VRG},  further gives
  \begin{eqnarray}\label{RG-V+1}
 \frac{\partial V_\Lambda G_0\lambda}{\partial \Lambda} 
 &=& \frac{\partial V_\Lambda}{\partial \Lambda} G_0\lambda+
V_\Lambda G_0 \frac{\partial \lambda}{\partial \Lambda} \nn
&=& V_\Lambda G_0 \frac{\partial \lambda}{\partial \Lambda} V_\Lambda G_0\lambda+
V_\Lambda G_0 \frac{\partial \lambda}{\partial \Lambda} 
\nn
&=& V_\Lambda G_0 \frac{\partial \lambda}{\partial \Lambda} (V_\Lambda G_0\lambda+1).
\end{eqnarray}
Using this in \eq{tmp} then gives the RG equation for the current operator, \eq{RGE}.

Our solution, \eq{our},  should be compared with the solution first given by NA in Ref.\ \cite{Nakamura:2006hc} and then used for RG 
 analyses in Refs.\ \cite{Nakamura:2006zp,Nakamura:2006hc,Nakamura:2007qm}:
 \begin{eqnarray}\label{nak}
\OL&=& \eta[1+V_{\bar\Lambda}\lambda G_0(1-V_{\bar\Lambda}G_0)^{-1}\lambda]\OLb
 [1+\lambda(1- G_0V_{\bar\Lambda})^{-1}G_0 \lambda V_{\bar\Lambda}]\eta 
\nn
 &=& \eta[1+V_{\bar\Lambda}\lambda (E'-H_0-V_{\bar\Lambda})^{-1}\lambda]\OLb
 [1+\lambda(E-H_0-  V_{\bar\Lambda})^{-1} \lambda V_{\bar\Lambda}]\eta .
 \end{eqnarray}
It is seen that our solution differs substantially from the one of Ref.\ \cite{Nakamura:2006hc}; in particular, the interaction operator  $V_{\bar\Lambda}$ in the denominators
 $(E'-H_0-V_{\bar\Lambda}\lambda)^{-1}$ and $(E-H_0- \lambda V_{\bar\Lambda})^{-1}$ of \eq{our}        is projected by $\lambda$, so that each intermediate state in the perturbation series for  $(E'-H_0-V_{\bar\Lambda}\lambda)^{-1}$ and $(E-H_0- \lambda V_{\bar\Lambda})^{-1}$ involves relative momenta restricted to the interval $\Lambda< k <\bar\Lambda$. By contrast, no such restriction on momenta appears in the corresponding intermediate states of \eq{nak}.

\section{Current operator RG equation from the cutoff invariance of an on-shell matrix element}

The RG equation for $\GamL$, \eq{RGE}, is derived in Ref.\ \cite{Nakamura:2006hc} as only a sufficient condition for the $\Lambda$ invariance of the physical matrix element of $\OL$: 
\be
\frac{\partial\la\OL\ra}{\partial\Lambda}=\frac{\partial}{\partial\Lambda}\bpsi_\B \eta\GamL(E_\B,E_\A)\eta\psi_\A=0 .
\eqn{naka}
\ee
That is,\footnote{To save on notation we suppress total momentum variables from the argument of $\GL$.}
\be
\la\OL\ra\equiv \bpsi_\B \eta\GamL(E_\B,E_\A)\eta\psi_\A
 = \bpsi_\B \Gamma^\mu(E_\B,E_\A)\psi_\A \eqn{our1}
\ee
where 
\be
\Gamma^\mu(E_\B,E_\A) \equiv O^\mu_{\bar\Lambda}(E_\B,E_\A) = \left. \eta\GamL(E_\B,E_\A)\eta\right|_{\Lambda=\bar\Lambda}
\ee
can be identified with the current vertex function of the full space \cite{Kvinikhidze:1998xn} in the limit $\bar\Lambda\rightarrow\infty$.
The sandwiching two-body wave functions $\bpsi_\B $ and $\psi_\A$ include bound states, and scattering states whose relative momenta, $p'$ and $p$, respectively, are smaller than the cutoff parameter: $p', p <\Lambda$. Although not emphasized in Ref.\ \cite{Nakamura:2006hc}, this restriction is essential for the derivation of \eq{RGE} in the case of scattering states for which the inhomogeneous term in
\be
\psie(\kk)=(2\pi)^3\delta^3(\p-\kk)+G_0(E_p,k)\int V_\Lambda(E_p;\kk,\kk') \theta(\Lambda-k')\psie(\kk')\frac{d^3k'}{(2\pi)^3}  \eqn{psi}
\ee
is not zero. In \eq{psi}, $E_{p'}$ and $E_p$ are on-shell energies obeying relations like $E_p=P^2/4m+p^2/m$. 
Indeed, retaining the inhomogeneous term of \eq{psi} and then substituting into \eq{naka} gives
\begin{align}
0\, &=\ \frac{\partial}{\partial\Lambda}\psiep \eta\GamL(E_{p'},E_p)\eta\psie=
\psiep \eta\GamL(E_{p'},E_p)\frac{\partial\eta}{\partial\Lambda}\psie+\dots
\nn[3mm]
&=\ 
\int \psiep(\kk')\frac{d^3k'}{(2\pi)^3} \theta(\Lambda-k')\GamL(E_{p'},E_p,\kk',\kk)\frac{\partial\theta(\Lambda-k)}{\partial\Lambda}
\frac{d^3k}{(2\pi)^3}
\nn
&\hspace{1cm} \times \left[(2\pi)^3\delta^3(\p-\kk)+\frac{1}{E_p-E_k}\int V_\Lambda(E_p;\kk,\kk'')\theta(\Lambda-k'')\psie(\kk'')\frac{d^3k''}{(2\pi)^3}\right]+\dots
\nn[3mm]
&
=
\int \psiep(\kk')\frac{d^3k'}{(2\pi)^3} \theta(\Lambda-k')\GamL(E_{p'},E_p,\kk',\p)
\frac{\partial\theta(\Lambda-p)}{\partial\Lambda}
\nn
&\hspace{1cm} +\int \psiep(\kk')\frac{d^3k'}{(2\pi)^3} \theta(\Lambda-k')\GamL(E_{p'},E_p,\kk',\kk)
\frac{\partial\theta(\Lambda-k)}{\partial\Lambda}
\frac{d^3k}{(2\pi)^3}\nn
&\hspace{2cm} \times\frac{1}{E_p-E_k} V_\Lambda(E_p;\kk,\kk'')\theta(\Lambda-k'')\psie(\kk'')\frac{d^3k''}{(2\pi)^3}+\dots
\nn[3mm]
&=\psiep \eta\GamL(E_{p'},E_p)\frac{\partial\eta}{\partial\Lambda}G_0(E_p)V_\Lambda(E_p)\eta\psie\nn
&\hspace{1cm}+
\psiep \eta V_\Lambda(E_{p'})G_0(E_{p'})\frac{\partial\eta}{\partial\Lambda}\GamL(E_{p'},E_p)\eta\psie
+\psiep \eta\frac{\partial\GamL(E_{p'},E_p)}{\partial\Lambda}\eta\psie
\eqn{in-hom}
\end{align}
where the last equality follows because we consider only low energy states with $p', p<\Lambda$ [in which case $\partial\theta(\Lambda-p')/\partial\Lambda=\partial\theta(\Lambda-p)/\partial\Lambda=0$]. Although this restriction is mentioned in Ref.\ \cite{Nakamura:2006hc}, we would like to stress here that this is not only a physical restriction, it also prevents \eq{our1} from a possible mathematical inconsistency of having more equations than the number of unknown variables. This point is clarified under  \eq{our2}.

\section{Current operator RG equation from the cutoff invariance of an off-shell five-point function }

The RG equation for the current operator, \eq{RGE}, is only a sufficient condition 
for the $\Lambda$ independence of $\bar\psi\eta \GL\eta\psi$, as expressed by
\eq{in-hom}, even though this equation involves matrix elements between {\em all} states $\psi$  (bound and scattering). Here we eliminate this ambiguity in the validity of \eq{RGE} by showing that this RG equation is a sufficient and {\it necessary} condition for $\Lambda$ independence of the five-point function $\eta G \eta\GamL\eta G \eta$. In other words, rather than basing the RG approach to the current operator on the condition [\eq{our1}]
\be
\bar\psi\eta \GL\eta\psi=\bar\psi\Gamma^\mu\psi
\ee
 for all $\Lambda<\bar\Lambda$, we suggest that it be based on the condition
\be
\eta G \eta\GamL\eta G \eta = \eta G^\mu\eta  \eqn{newdef}
\ee
 for all $\Lambda<\bar\Lambda$,  where $G^\mu$ is the five-point function defined  as
\be
G^\mu = \left. G \eta\GamL\eta G \right|_{\bar\Lambda=\Lambda} =G \Gamma^\mu G.
\ee
We note that $G^\mu$
corresponds to the two-body Green function $G$ with all possible insertions of a current \cite{Kvinikhidze:1998xn}. In the five-point function $\eta G^\mu\eta$, 
 neither the incoming nor outgoing two-body states are on the energy shell; by contrast, both these states are on the energy shell in $\bar\psi\Gamma^\mu\psi$. At the same time, such five-point Green functions are necessary ingredients for three-body currents where two-body subsystems are off shell. In this sense the use of $\eta G^\mu\eta$ for the RG approach to the current operator, is naturally related to the RG approach to the two-body interaction, where the cutoff independence of the fully off-shell two-body scattering amplitude is used \cite{Birse:1998dk}. \eq{newdef} thus defines the effective current vertex $\GamL$ so that the five point Green function $G \eta\GamL\eta G $ coincides with the  five point Green function $G^\mu=G\Gamma^\mu G$  if the relative momenta of incoming and outgoing nucleons are below $\Lambda$. Showing the two-body energy arguments, \eq{newdef} is
 \begin{align}\label{our2}
 \eta G(E') \eta &\GamL(E',E)\eta G(E) \eta =\eta G^\mu(E',E) \eta\nn
&=\eta G(E')\Gamma^\mu(E',E)G(E)\eta
\end{align}
where the external $\eta$'s ensure the above mentioned restriction on the relative momenta of incoming and outgoing particles. Without this restriction, one would have the self-consistency constraint  $\Gamma^\mu= \eta\GamL\eta$, which would simply mean that the model is cutoff by $\Lambda$ from the very beginning, leaving us with nothing further to be done.

The $\Lambda$ independence specified by \eq{our2}  leads to the RG equation
\begin{align}
&%
\eta G(E')\frac{\partial   \eta\GamL(E',E)\eta   }{\partial \Lambda}G(E)\eta=0  .
\eqn{stronger}
\end{align}

The cutoff-independence condition of \eq{our1} differs from the one of \eq{our2}. It is also a weaker condition as \eq{our1} follows from \eq{our2}; i.e.,  \eq{our1} is necessary but not sufficient condition for \eq{our2} to be satisfied. The  essential difference between these two conditions is that  \eq{our2}   involves off-shell scattering amplitudes through $\eta G(E) \eta$, whereas \eq{our1} involves half-on-shell amplitudes through the scattering states
\be
\psie(\kk)=(2\pi)^3\delta(\p-\kk)+G_0(E_p,k)T(E_p;\kk,\p) .
\ee
In this sense \eq{our1} looks more like an extension of the RG approach discussed in Ref.\ \cite{Bogner:2001gq} (which is based on the independence of the {\it half-on-shell} scattering amplitude) to the case of current operators. Moreover,
\eq{our2} defines $\GamL(E',E)$ {\it unambiguously} whereas there is an ambiguity in its definition by \eq{our1} (this ambiguity was already poined out in Ref.\  \cite{Nakamura:2006hc}). 

To see the difference between these two conditions yet more precisely, let's write  \eq{naka} in expanded form for the case of scattering states: 
 \begin{align}\label{naka1}
\frac{\partial}{\partial\Lambda}&
\int \left[(2\pi)^3\delta(\p'-\kk')+T(E_{p'};\p',\kk')G_0(E_{p'},k') \right]
\frac{d^3k'}{(2\pi)^3} \theta(\Lambda-k')\GamL(E_{p'},E_p,\kk',\kk)
\nn
&
\theta(\Lambda-k)
\frac{d^3k}{(2\pi)^3}
\left[(2\pi)^3\delta(\kk-\p)+G_0(E_p,k)T(E_p;\kk,\p) \right] = 0,
\end{align}
and likewise \eq{stronger} after dividing out external $G_0$ factors:
 \begin{align}\label{stronger1}
\frac{\partial}{\partial\Lambda}&
\int \left[(2\pi)^3\delta(\p'-\kk')+T(E';\p',\kk')G_0(E',k') \right]
\frac{d^3k'}{(2\pi)^3} \theta(\Lambda-k')\GamL(E',E,\kk',\kk)
\nn
&
\theta(\Lambda-k) \frac{d^3k}{(2\pi)^3} \left[(2\pi)^3\delta(\kk-\p)+G_0(E,k)T(E;\kk,\p) \right] = 0
\end{align}
where no restriction is put on $E'$ and $E$. It is now clear that only that part of \eqs{stronger1} corresponding to $E=E_p$, and $E'=E_{p'}$ reproduces the whole set of \eqs{naka1}. With no restriction being put on the external relative momenta $\p$ and $\p'$ of \eq{stronger1} (apart from $p', p<\Lambda$), one can invert the external Green functions to obtain the RG equation for $\GL$.
That is why the RG equation for $\GamL$ is not only a sufficient but also a {\it  necessary} condition for  \eq{stronger}, whereas it is only a sufficient condition for \eq{naka}.

 To show explicitly how one obtains the RG equation unambiguously, we use \eqs{useful} and the shorthand notation $\delta\equiv \partial\eta/\partial\Lambda=-\partial\lambda/\partial\Lambda$ in the following:
  \begin{align} \label{our-RG1}
 0& =\eta G\frac{\partial   \eta\GamL\eta   }{\partial \Lambda}G\eta
 =\eta G\left(\eta   \frac{\partial \GamL}{\partial \Lambda}\eta   +\delta\GamL\eta+\eta\GamL\delta\right)G\eta
 \nn
 &=\eta G\eta   \frac{\partial \GamL}{\partial \Lambda}\eta G\eta   +\eta G\delta\GamL\eta G\eta+\eta G\eta\GamL\delta G\eta
 \nn
 &
 =\eta G\eta   \frac{\partial \GamL}{\partial \Lambda}\eta G\eta   +(\eta G_0+ \eta G \eta V_\Lambda G_0)\delta\GamL\eta G\eta+\eta G\eta\GamL\delta (\eta G_0+ G_0V_\Lambda\eta G \eta )
 \nn
 &
 =\eta G\eta   \frac{\partial \GamL}{\partial \Lambda}\eta G\eta   + \eta G \eta V_\Lambda G_0\delta\GamL\eta G\eta+\eta G\eta\GamL\delta G_0V_\Lambda\eta G \eta 
 +\eta G_0\delta\GamL\eta G\eta+\eta G\eta\GamL\delta \eta G_0
  \nn
 &
 =\eta G\eta\left(\frac{\partial \GamL}{\partial \Lambda}+ V_\Lambda G_0\delta\GamL+\GamL\delta G_0V_\Lambda \right)\eta G \eta 
 +\eta G_0\delta\GamL\eta G\eta+\eta G\eta\GamL\delta \eta G_0 .
 \end{align}
Furthermore, as we are interested in external relative momenta strictly below $\Lambda$, the last two terms of \eq{our-RG1} are zero since $\delta \eta=0$.
One can then invert $\eta G \eta$
 in the reduced subspace by acting on \eq{our-RG1} with $1-V_\Lambda G_0\eta$ from the right side and with $1-\eta G_0 V_\Lambda $ from the left:
 \begin{eqnarray} \label{our-RG2}
 0&=&(1-\eta G_0 V_\Lambda )\eta G\eta\left(  \frac{\partial \GamL}{\partial \Lambda}+ 
 V_\Lambda G_0\delta\GamL+\GamL\delta G_0V_\Lambda \right)\eta G \eta (1-V_\Lambda G_0\eta)
 \nn
 &=& G_0\eta\left(  \frac{\partial \GamL}{\partial \Lambda}+ 
 V_\Lambda G_0\delta\GamL+\GamL\delta G_0V_\Lambda \right)\eta G_0 \eqn{end}
 \end{eqnarray}
 where \eqs{invert} have been used. In this way we derive the RG equation for the current operator, \eq{RGE}, unambiguously.
 
 \section{$\GamL$ directly from its definition}

In Section II we proposed \eq{our} as the solution to the RG equation, and proved it by direct substitution into this equation. Here we make doubly sure that no subtleties were missed in the process, by deriving \eq{our} directly 
from its definition, \eq{our2}.

Starting with \eq{our2}, we can use \eqs{der-cut} to write
\begin{align}
\eta G \eta\GamL\eta G \eta 
&= \eta G\Gamma^\mu G \eta \nn
&=\eta G \eta (1-V_{\bar\Lambda} G_0\lambda)^{-1}\Gamma^\mu
 (1-\lambda G_0 V_{\bar\Lambda} )^{-1}\eta G\eta .
\end{align}
The $\eta G\eta$ factors can then be removed, as in \eq{end},  by acting with $1-V_\Lambda G_0\eta$ from the right and  $1-\eta G_0 V_\Lambda $ from the left, and then using \eqs{invert}:
\be
   \eta\GamL\eta = \eta (1-V_{\bar\Lambda} G_0\lambda)^{-1}\Gamma^\mu 
  (1-\lambda G_0 V_{\bar\Lambda} )^{-1}\eta 
  \ee
 which is just what we proposed in \eq{our}.
  
\section{Current conservation}

The question of how to properly implement current conservation in EFT with a cutoff, so that
gauge invariance is ensured in practical calculations, is a subtle one \cite{prep}.  
Here we show that the problem of current conservation in the RG approach is likewise not so simple (it is certainly not as simple as presented in Ref.\ \cite{Nakamura:2006hc}).

In order to avoid the well known problems of current conservation in theories with a finite cutoff, we consider the simple  case where the starting cutoff  is taken to infinity,
$\bar\Lambda=\infty$. Then in the best case we will have the usual two-body Ward-Takahashi (WT) identities \cite{Bentz:1986nq}
\begin{subequations}  \eqn{wt}
\begin{align}
q_\mu G^\mu(E',E) &=\Gamma_0^{0}G(E)-G(E')\Gamma_0^{0}  , \eqn{wta} \\
q_\mu\Gamma^\mu(E',E) &=G^{-1}(E')\Gamma_0^{0}-\Gamma_0^{0}G^{-1}(E) \eqn{wtb}
\end{align}
\end{subequations}
where $\Gamma_0^0$ is zeroth component of the current operator $\Gamma_0^\mu$ of two non-interacting particles, and is specified for initial (final) total four-momentum $P=p_1+p_2$ ($P'=p'_1+p'_2$) and relative momentm $\p$ ($\p'$) as
\begin{eqnarray}
\la \p'|\Gamma_0^{0}(P',P)|\p\ra&=&i(2\pi)^3\left[e_1 \delta(\p'_2-\p_2)+e_2\delta(\p'_1-\p_1)\right]
\nn
&=&i(2\pi)^3\left[e_1\delta({\bf p}'-{\bf p}-{\bf q}/2)+e_2\delta({\bf p}'-{\bf p}+{\bf q}/2)\right]
\end{eqnarray}
It is important to realize that the WT identities of \eqs{wt} are damaged after the introduction of a finite momentum cutoff $\Lambda$. In particular, introducing the cutoff into \eq{wta} gives
\begin{align}
q_\mu \eta G^\mu(E',E) \eta &= \eta \Gamma_0^{0}G(E) \eta- \eta G(E')\Gamma_0^{0} \eta  .
\eqn{wtca}
\end{align}
Because $\eta \Gamma_0^{0}G(E) \eta\neq\eta \Gamma_0^{0}\eta G(E) \eta$ (the cutoff $\eta$ does not commute with $\Gamma_0^{0}$), it is evident that \eq{wtca} is not of the same form as \eq{wta}, so it is not a usual WT identity. Similarly, to see how the WT identity for current operator $\OL=\eta  \GamL \eta$ (which does depend on $\Lambda$) differs from the usual  one, we use \eq{our} and \eq{wtb} to write
\begin{align}
q_\mu \eta  \GamL(E',E) \eta &=\eta [1-V_{\bar\Lambda} G_0(E')\lambda]^{-1}\left[
G^{-1}(E')\Gamma_0^{0}-\Gamma_0^{0}G^{-1}(E)\right]\nn
&\hspace{4mm} \times  [1-\lambda G_0(E) V_{\bar\Lambda} ]^{-1}\eta \nn[3mm]
 &=\eta [1+V_{\Lambda} G_0(E')\lambda]\left[
G^{-1}(E')\Gamma_0^{0}-\Gamma_0^{0}G^{-1}(E)\right]\nn
&\hspace{4mm} \times  [1+\lambda G_0(E) V_{\Lambda} ]\eta .
\end{align}
This expression can be simplified using 
\begin{subequations}
\begin{align}
G^{-1}(1+\lambda G_0 V_\Lambda) &= G_0^{-1} - \eta V_\Lambda \\
(1+ V_\Lambda  G_0 \lambda) G^{-1}&= G_0^{-1} - V_\Lambda  \eta
\end{align}
\end{subequations}
which follow from the equations for $G$, \eqs{greena} and (\ref{greenb}). Thus
\begin{align}
q_\mu \eta  \GamL(E',E) \eta &= \eta [G_0^{-1}(E')-V_\Lambda (E')]\eta \Gamma_0^{0}
 [1+\lambda G_0(E)V_\Lambda (E)] \eta
 \nn
 &
 - \eta [1+V_\Lambda (E')G_0(E')\lambda]\Gamma_0^{0} \eta [G_0^{-1}(E)-V_\Lambda (E)]\eta ,
\end{align}
which can be written safely without the energy arguments:
\begin{align}
q_\mu \eta  \GamL\eta &=
\eta \left\{G_0^{-1}-V_\Lambda \right\}\eta \Gamma_0^{0}
 (1+\lambda G_0V_\Lambda ) \eta
 - \eta (1+V_\Lambda G_0\lambda)\Gamma_0^{0} \eta \left\{G_0^{-1}-V_\Lambda \right\}\eta 
 \eqn{wtcb}
 \end{align}
 Although \eq{wtcb} is not a usual WT identity, it still leads to a conserved current
due to the operators  in the curly brackets, $\left\{G_0^{-1}-V_\Lambda \right\}$:
\be
q_\mu \psiep\eta  \GamL(E',E)\eta\psie=0 .
\ee
Having derived the modified WT identity, 
\eq{wtcb}, it is easy to realize that there was no obligation of pushing the starting cutoff to infinity. We could have started with a  finite cutoff $\bar\Lambda$; however, our starting
WT identities would then need to be \eqs{wtca} and (\ref{wtcb}) (with $\Lambda$ replaced by $\bar\Lambda$), instead of the usual ones, \eqs{wt}. In this way we would come to the same result [\eqs{wtca} and (\ref{wtcb}) for any $\Lambda<\bar\Lambda$].

It is important to note that the modified WT identity, \eq{wtcb},  relates the reduced space effective current vertex $\GL$,  only to the corresponding effective potential $V_\Lambda$ ($V_{\bar\Lambda}$ is not involved), and that it enters the WT identity only with relative momenta below $\Lambda$ for all physically interesting low energy transitions.
These properties are indispensible for constructing a self-contained EFT in the reduced momentum space \cite{prep}.

\begin{acknowledgments}

The research described in this publication was made possible in part by GNSF Grant No.\ GNSF/ST06/4-050.\end{acknowledgments}

\appendix*
\section{Useful equations}
Here we gather together some standard equations of the RG approach that are made use of in the main text. 
Firstly, we note that the solution of the RG equation for the effective potential, \eq{VRG},  can be formally written in terms of the initial potential  $V_{\bar\Lambda}$ as
\begin{subequations}
\begin{align}
V_\Lambda &=(1-V_{\bar\Lambda}G_0\lambda)^{-1}V_{\bar\Lambda},  \eqn{VLa} \\
 &=V_{\bar\Lambda}(1-\lambda G_0 V_{\bar\Lambda})^{-1}.  \eqn{VLb}
\end{align} \eqn{VL}
\end{subequations}
These equations then give the useful relations
\begin{subequations}
\begin{align}
(1-V_{\bar\Lambda}G_0\lambda)^{-1} &=1+V_{\Lambda}G_0\lambda,   \\
 (1-\lambda G_0 V_{\bar\Lambda})^{-1}&=1+\lambda G_0 V_{\Lambda}.  
\end{align} \eqn{VLP}
\end{subequations}
Secondly, we note that in this paper we assume that a finite value of $\bar\Lambda$ defines the full model space. That is, all relative momenta are assumed to lie within a sphere of radius $\bar\Lambda$ so that
\be
\bar\eta \equiv \left. \eta\right|_{\Lambda=\bar\Lambda} = 1, \hspace{1cm}
\bar\lambda\equiv \left. \lambda \right|_{\Lambda=\bar\Lambda} = 0,
\ee
and
\be
\lambda = 1 - \eta.
\ee

We also note that \eq{LSE}, and its reverse form $T=V_\Lambda+T G_0\eta V_\Lambda$, imply that the two-body Green function $G\equiv G_0+G_0 T G_0$ satisfies the equations
\begin{subequations}
\begin{align}
G&=G_0+(\lambda G_0+G\eta ) V_\Lambda G_0 , \eqn{greena} \\
G&=G_0+ G_0V_\Lambda(\lambda G_0+\eta G) ,  \eqn{greenb} \\
G&=G_0+ G  V_{\bar\Lambda} G_0 ,   \eqn{greenc}\\
G&=G_0+ G_0V_{\bar\Lambda}G .  \eqn{greend}
\end{align}
\end{subequations}
These then imply the following equations (together with their reversed forms):
\begin{subequations} \eqn{useful}
\begin{align}
\eta G &=\eta G_0+\eta G \eta V_\Lambda G_0, \eqn{eq-tGt} \\
\eta G &=\eta G_0+\eta G V_{\bar\Lambda} G_0 ,  \eqn{usefulc}\\
\eta G & \eta=\eta G_0+\eta G V_{\bar\Lambda} G_0\eta ,
\end{align}
\end{subequations}
The first of these equations, and its reversed form, then give
\begin{subequations}   \eqn{invert}
\begin{align}
\eta G \eta(1-V_\Lambda G_0\eta)&=\eta G_0, \\
(1-\eta G_0V_\Lambda )\eta G \eta&= G_0\eta, 
 \end{align}
\end{subequations}
while the last two equations of \eqs{useful}, and their reversed forms, give
\begin{subequations}   \eqn{der-cut}
\begin{align}
\eta G &= \eta G  \eta  \left(1-  V_{\bar\Lambda} G_0\lambda \right)^{-1} , \\
 G \eta&=   \left(1- \lambda G_0 V_{\bar\Lambda} \right)^{-1} \eta G  \eta.
\end{align}
\end{subequations}


\begin{thebibliography}{16}
\expandafter\ifx\csname natexlab\endcsname\relax\def\natexlab#1{#1}\fi
\expandafter\ifx\csname bibnamefont\endcsname\relax
  \def\bibnamefont#1{#1}\fi
\expandafter\ifx\csname bibfnamefont\endcsname\relax
  \def\bibfnamefont#1{#1}\fi
\expandafter\ifx\csname citenamefont\endcsname\relax
  \def\citenamefont#1{#1}\fi
\expandafter\ifx\csname url\endcsname\relax
  \def\url#1{\texttt{#1}}\fi
\expandafter\ifx\csname urlprefix\endcsname\relax\def\urlprefix{URL }\fi
\providecommand{\bibinfo}[2]{#2}
\providecommand{\eprint}[2][]{\url{#2}}

\bibitem[{\citenamefont{Wilson and Kogut}(1974)}]{Wilson:1973jj}
\bibinfo{author}{\bibfnamefont{K.~G.} \bibnamefont{Wilson}} \bibnamefont{and}
  \bibinfo{author}{\bibfnamefont{J.~B.} \bibnamefont{Kogut}},
  \bibinfo{journal}{Phys. Rept.} \textbf{\bibinfo{volume}{12}},
  \bibinfo{pages}{75} (\bibinfo{year}{1974}).

\bibitem[{\citenamefont{Polchinski}(1984)}]{Polchinski:1983gv}
\bibinfo{author}{\bibfnamefont{J.}~\bibnamefont{Polchinski}},
  \bibinfo{journal}{Nucl. Phys.} \textbf{\bibinfo{volume}{B231}},
  \bibinfo{pages}{269} (\bibinfo{year}{1984}).

\bibitem[{\citenamefont{Morris}(1998)}]{Morris:1998da}
\bibinfo{author}{\bibfnamefont{T.~R.} \bibnamefont{Morris}},
  \bibinfo{journal}{Prog. Theor. Phys. Suppl.} \textbf{\bibinfo{volume}{131}},
  \bibinfo{pages}{395} (\bibinfo{year}{1998}), \eprint{hep-th/9802039}.

\bibitem[{\citenamefont{Birse et~al.}(1999)\citenamefont{Birse, McGovern, and
  Richardson}}]{Birse:1998dk}
\bibinfo{author}{\bibfnamefont{M.~C.} \bibnamefont{Birse}},
  \bibinfo{author}{\bibfnamefont{J.~A.} \bibnamefont{McGovern}},
  \bibnamefont{and} \bibinfo{author}{\bibfnamefont{K.~G.}
  \bibnamefont{Richardson}}, \bibinfo{journal}{Phys. Lett.}
  \textbf{\bibinfo{volume}{B464}}, \bibinfo{pages}{169} (\bibinfo{year}{1999}),
  \eprint{hep-ph/9807302}.

\bibitem[{\citenamefont{Birse}(2006)}]{Birse:2005um}
\bibinfo{author}{\bibfnamefont{M.~C.} \bibnamefont{Birse}},
  \bibinfo{journal}{Phys. Rev. C} \textbf{\bibinfo{volume}{74}},
  \bibinfo{pages}{014003} (\bibinfo{year}{2006}), \eprint{nucl-th/0507077}.

\bibitem[{\citenamefont{Harada et~al.}(2006)\citenamefont{Harada, Inoue, and
  Kubo}}]{Harada:2005tw}
\bibinfo{author}{\bibfnamefont{K.}~\bibnamefont{Harada}},
  \bibinfo{author}{\bibfnamefont{K.}~\bibnamefont{Inoue}}, \bibnamefont{and}
  \bibinfo{author}{\bibfnamefont{H.}~\bibnamefont{Kubo}},
  \bibinfo{journal}{Phys. Lett.} \textbf{\bibinfo{volume}{B636}},
  \bibinfo{pages}{305} (\bibinfo{year}{2006}), \eprint{nucl-th/0511020}.

\bibitem[{\citenamefont{Harada and Kubo}(2006)}]{Harada:2006cw}
\bibinfo{author}{\bibfnamefont{K.}~\bibnamefont{Harada}} \bibnamefont{and}
  \bibinfo{author}{\bibfnamefont{H.}~\bibnamefont{Kubo}},
  \bibinfo{journal}{Nucl. Phys.} \textbf{\bibinfo{volume}{B758}},
  \bibinfo{pages}{304} (\bibinfo{year}{2006}), \eprint{nucl-th/0605004}.

\bibitem[{\citenamefont{Harada et~al.}(2007)\citenamefont{Harada, Kubo, and
  Ninomiya}}]{Harada:2007ua}
\bibinfo{author}{\bibfnamefont{K.}~\bibnamefont{Harada}},
  \bibinfo{author}{\bibfnamefont{H.}~\bibnamefont{Kubo}}, \bibnamefont{and}
  \bibinfo{author}{\bibfnamefont{A.}~\bibnamefont{Ninomiya}}
  (\bibinfo{year}{2007}), \eprint{nucl-th/0702074}.

\bibitem[{\citenamefont{Nakamura}(2005)}]{Nakamura:2004ek}
\bibinfo{author}{\bibfnamefont{S.~X.} \bibnamefont{Nakamura}},
  \bibinfo{journal}{Prog. Theor. Phys.} \textbf{\bibinfo{volume}{114}},
  \bibinfo{pages}{77} (\bibinfo{year}{2005}), \eprint{nucl-th/0411108}.

\bibitem[{\citenamefont{Nakamura and
  Ando}(2006{\natexlab{a}})}]{Nakamura:2006hc}
\bibinfo{author}{\bibfnamefont{S.~X.} \bibnamefont{Nakamura}} \bibnamefont{and}
  \bibinfo{author}{\bibfnamefont{S.-i.} \bibnamefont{Ando}},
  \bibinfo{journal}{Phys. Rev. C} \textbf{\bibinfo{volume}{74}},
  \bibinfo{pages}{034004} (\bibinfo{year}{2006}{\natexlab{a}}),
  \eprint{nucl-th/0606026}.

\bibitem[{\citenamefont{Bentz}(1985)}]{Bentz:1986nq}
\bibinfo{author}{\bibfnamefont{W.}~\bibnamefont{Bentz}},
  \bibinfo{journal}{Nucl. Phys.} \textbf{\bibinfo{volume}{A446}},
  \bibinfo{pages}{678} (\bibinfo{year}{1985}).

\bibitem[{pre()}]{prep}
\bibinfo{note}{In preparation}.

\bibitem[{\citenamefont{Nakamura and
  Ando}(2006{\natexlab{b}})}]{Nakamura:2006zp}
\bibinfo{author}{\bibfnamefont{S.~X.} \bibnamefont{Nakamura}} \bibnamefont{and}
  \bibinfo{author}{\bibfnamefont{S.-i.} \bibnamefont{Ando}}
  (\bibinfo{year}{2006}{\natexlab{b}}), \eprint{nucl-th/0611052}.

\bibitem[{\citenamefont{Nakamura and G\aa{}rdestig}(2007)}]{Nakamura:2007qm}
\bibinfo{author}{\bibfnamefont{S.~X.} \bibnamefont{Nakamura}} \bibnamefont{and}
  \bibinfo{author}{\bibfnamefont{A.}~\bibnamefont{G\aa{}rdestig}}
  (\bibinfo{year}{2007}), \eprint{arXiv:0704.3757 [nucl-th]}.

\bibitem[{\citenamefont{Kvinikhidze and
  Blankleider}(1999)}]{Kvinikhidze:1998xn}
\bibinfo{author}{\bibfnamefont{A.~N.} \bibnamefont{Kvinikhidze}}
  \bibnamefont{and}
  \bibinfo{author}{\bibfnamefont{B.}~\bibnamefont{Blankleider}},
  \bibinfo{journal}{Phys. Rev. C} \textbf{\bibinfo{volume}{60}},
  \bibinfo{pages}{044003} (\bibinfo{year}{1999}), \eprint{nucl-th/9901001}.

\bibitem[{\citenamefont{Bogner et~al.}(2003)\citenamefont{Bogner, Kuo, Schwenk,
  Entem, and Machleidt}}]{Bogner:2001gq}
\bibinfo{author}{\bibfnamefont{S.~K.} \bibnamefont{Bogner}},
  \bibinfo{author}{\bibfnamefont{T.~T.~S.} \bibnamefont{Kuo}},
  \bibinfo{author}{\bibfnamefont{A.}~\bibnamefont{Schwenk}},
  \bibinfo{author}{\bibfnamefont{D.~R.} \bibnamefont{Entem}}, \bibnamefont{and}
  \bibinfo{author}{\bibfnamefont{R.}~\bibnamefont{Machleidt}},
  \bibinfo{journal}{Phys. Lett.} \textbf{\bibinfo{volume}{B576}},
  \bibinfo{pages}{265} (\bibinfo{year}{2003}), \eprint{nucl-th/0108041}.

\end{thebibliography}

\end{document}